# Shape of spin density wave versus temperature in $AFe_2As_2$ (A=Ca, Ba, Eu): A Mössbauer study

A. Błachowski [1], K. Ruebenbauer [1*], J. Żukrowski [2], K. Rogacki [3], Z. Bukowski [4], and J. Karpinski [4]

[1] Mössbauer Spectroscopy Division, Institute of Physics, Pedagogical University
PL-30-084 Kraków, ul. Podchorążych 2, Poland

[2] Solid State Physics Department, Faculty of Physics and Applied Computer Science, AGH University of Science and Technology
PL-30-059 Kraków, Al. Mickiewicza 30, Poland

[3] Institute of Low Temperatures and Structure Research, Polish Academy of Sciences
PL-50-422 Wrocław, ul. Okólna 2, Poland

[4] Laboratory for Solid State Physics, ETH Zurich
CH-8093 Zurich, Switzerland

[*] Corresponding author: sfrueben@cyf-kr.edu.pl

**PACS:** 74.70.Xa, 75.30.Fv, 76.80.+y

**Keywords:** iron-based superconductors, spin density waves, Mössbauer spectroscopy

## Abstract

Parent compounds $AFe_2As_2$ (A=Ca, Ba, Eu) of the 122 family of the iron-based superconductors have been studied by $^{57}Fe$ Mössbauer spectroscopy in the temperature range 4.2 K - ~300 K. Spin density waves (SDW) have been found with some confidence. They are either incommensurate with the lattice period or the ratio of the respective periods is far away from ratio of small integers. SDW shape is very unconventional i.e. differs from the sinusoidal shape. Magnetic order starts with lowered temperature as narrow sheets of the significant electron spin density separated by areas with very small spin density. Magnetic sheets are likely to be ordered in the alternate anti-ferromagnetic fashion as the material as whole behaves similarly to collinear anti-ferromagnet. Further lowering of temperature simply expands sheet thickness leading to the near triangular SDW. Finally, sheets fill the whole available space and almost rectangular shape of SDW is reached. Substantial maximum amplitude of SDW appears at the temperature just below the magnetic onset temperature, and this maximum amplitude increases slightly with lowering temperature. The square root from the mean squared hyperfine field behaves versus temperature according to the universality class $(1,2)$, i.e., with the electronic spin space having dimensionality equal unity and the real space having dimensionality equal two. The more or less pronounced tail above transition temperature due to the development of incoherent SDW is seen.

## 1. Introduction

Iron pnictides and chalcogenides are very interesting compounds as by slight variation of the atomic order and lattice parameters due to doping or applying external pressure one can switch from anti-ferromagnets to the iron-based superconducting materials. The material in the metallic state could be either superconductor at low temperatures without any 3d magnetism [1, 2] or some magnetic order develops for delocalized spins of the 3d electrons [1]. Iron-based pnictides $AFe_2As_2$ (A=Ca, Sr, Ba, Eu) are parent compounds of the 122 family of the iron-based superconductors. The parents are strongly layered intermetallic compounds developing iron-based magnetic order in contrast to corresponding superconductors, the latter having no magnetic moment due to iron [3]. A magnetic transition is correlated with the phase transition from the high temperature tetragonal phase to the low temperature orthorhombic phase. Both transitions occur at about 170 K for A=Ca [4], 200 K for Sr [5], 140 K for Ba [6] and 190 K for Eu [5]. A magnetic order in the parent compounds exhibits many unconventional features typical for the spin density wave (SDW) type of order. Other studies detected SDW, in fact, at least for under-doped materials [7, 8]. However, many studies by methods sensitive to the local environment of the iron atom refuted existence of SDW being incommensurate with the lattice period [9, 10]. Bonville *et al.* [11] have found SDW in Co-substituted (under-doped) $BaFe_2As_2$ by using Mössbauer spectroscopy. The shape of SDW was found to be far away from the sinusoidal in contrast to the “classical” SDW materials like chromium. Some other attempts to evaluate Mössbauer data within framework of SDW have been performed as well, but the assumption of the purely sinusoidal SDW shape was applied in the more or less implicit fashion [12]. Mössbauer spectra of the parent compounds of the 122 and 1111 families of iron-based superconductors in the magnetically ordered state (even close to saturation) are hard to fit by unique set of hyperfine parameters, despite fact that iron occupies unique and precisely defined crystallographic position including black and white symmetry [13-17]. The black and white symmetry is due to the anti-ferromagnetism. Better fits are usually obtained applying field distribution or several fields, but such treatment has no physical background. In order to resolve this discrepancy we have performed studies of the $AFe_2As_2$ (A=Ca, Ba, Eu) compounds in the temperature range 4.2 K - ~300 K by using Mössbauer spectroscopy of the 14.41-keV line in $^{57}Fe$. One has to note that above compounds make the best ordered crystals of all parents of the iron-based superconductors as their atomic composition ensures that the number of point defects is negligible.

## 2. Experimental

Single crystals of $CaFe_2As_2$ and $EuFe_2As_2$ were grown by the tin flux method as described in Ref. [18] and [19], respectively. The polycrystalline sample of $BaFe_2As_2$ was prepared by the solid state reaction at elevated temperature [20]. X-ray diffraction confirmed that samples are single phases of high purity [18-20].

Mössbauer absorbers were prepared in the powder form crushing single crystals or polycrystalline material in the mortar to the micrometer size crystals. Powder absorbers are advantageous, while seeking for details of the hyperfine parameter distributions unless perfect single crystals filling completely absorber holder and sufficiently thin are available, as partial orientation interferes with the hyperfine field distribution. On the other hand, hyperfine parameters (line positions) do not depend on the crystal size unless this crystal has size comparable with the chemical unit cell. One cannot expect very large coherence length of SDW due to pinning by defects and therefore micrometer scale of the powder grain is large compared to the average distance between defects. Powders were mixed with the ample amount of boron carbide fine powder to assure as random as possible orientation of crystallites and lightly pressed between mylar sheets aluminized on both sides each. About 55 mg of $AFe_2As_2$ material was used to make absorber, the latter having 16 mm diameter with 12 mm clear bore. Absorbers were mounted on the gold plated copper holder equipped with the thermometer and heater. Janis Research Co. Inc. cryostat SVT-400M was used with the sample either immersed in the liquid helium under almost ambient pressure or exposed to the nitrogen vapor flow. For the latter case the heater was used to maintain desired temperature. The long range temperature stability and homogeneity was better than 0.01 K except for room temperature (RT) and 4.2 K, where the temperature was not stabilized. Commercial source $^{57}$Co(Rh) from Ritverc G.m.b.H. was attached to the transducer and maintained at ambient temperature. Spectra were collected by means of the RENON MsAa-3 spectrometer with the velocity scale calibrated by the Michelson-Morley interferometer equipped with the metrological quality He-Ne laser. Calibration data were corrected for the beam divergence. The round-corner triangular mode of the spectrometer operation was applied. LND Inc. Kr-filled proportional detector with iron-free beryllium window was used to detect γ-rays. All shifts of the spectra are reported here versus room temperature α-Fe. No signal from iron located in other phases than the investigated compound was found for all samples used.

## 3. Mössbauer spectra evaluation within SDW model

SDW represents stationary periodic field of the electron spin density within the crystal lattice. It is assumed that the hyperfine magnetic (induction) field seen by the atomic nucleus at some (average) position $\mathbf{r}$ of the crystal lattice is proportional to the SDW amplitude at the same position (some constant angle between SDW amplitude and hyperfine field cannot be excluded). Proportionality constant (usually) strongly depends on the kind of atom due to the atomic shell polarization. Hence, the problem could be considered as time independent. In general, one can express amplitude of the SDW (hyperfine field $\mathbf{B}$) by the following expression:

$$\mathbf{B}(\mathbf{r}) = \sum_{hkl=0}^{HKL} \mathbf{b}_{hkl} \cos\left[ hq_a \mathrm{x} + kq_b \mathrm{y} + lq_c \mathrm{z} + \Phi_{hkl} \right] \text{ with } \mathbf{r} = \begin{pmatrix} \mathrm{x} \\ \mathrm{y} \\ \mathrm{z} \end{pmatrix}. \tag{1}$$

Here, the indices $hkl$ stand for the Miller indices of the crystal lattice considered, while symbols $HKL$ denote maximum values of the respective Miller indices. The symbol $\mathbf{b}_{hkl}$ stands for the amplitude of the particular hyperfine field component due to the Miller indices $hkl$, while symbols $q_a$, $q_b$ and $q_c$ stand for the SDW wave-vector components along principal crystal axes $\mathbf{a}, \mathbf{b}$ and $\mathbf{c}$, respectively. The symbol $\Phi_{hkl}$ denotes relative phase angle of the respective SDW component, the latter having definite Miller indices set $hkl$. Co-ordinates $\mathrm{xyz}$ are defined in the principal crystal axes system $\mathbf{a}, \mathbf{b}$ and $\mathbf{c}$, respectively. Average atomic (nuclear) positions within crystal lattice are expressed as $(n_a + u)\mathrm{a}, (n_b + v)\mathrm{b}$ and $(n_c + w)\mathrm{c}$ along respective crystal axes. Symbols $n_a, n_b, n_c$ denote relative position of the particular chemical unit cell, while symbols $u, v, w$ denote fractal co-ordinates of the particular atom (site) within cell. Symbols $\mathrm{a, b, c}$ stand for respective lattice constants. SDW is commensurate provided $(2\pi)^{-1} q_a \mathrm{a}, (2\pi)^{-1} q_b \mathrm{b}$ and $(2\pi)^{-1} q_c \mathrm{c}$ are rational numbers. Otherwise, SDW is incommensurate. It is hard to distinguish these two behaviors for above rational numbers being far away from the (irreducible) ratio of small non-zero integers. In order to get reliable results some simplifications are usually necessary. It has been assumed that resonant atom occupies unique position within chemical unit cell. Furthermore, it has been assumed that SDW is collinear and of the (normal) anti-ferromagnetic character. The last assumption allows for replacement of the axial vector $\mathbf{B}$ by corresponding pseudo-scalar $B$ remembering that sign of the hyperfine field is inaccessible in the experimental arrangement used. Assumption about collinear anti-ferromagnetic behavior leads to another simplification allowing replacement of three Miller indices by single index $n$ and leading to replacement of vectors $\mathbf{b}_{hkl}$ by pseudo-scalars. Finally, it has been assumed that SDW has similar symmetry like previously observed SDW in e.g. chromium. The latter SDW could be expanded in the sine functions (SDW has definite parity) with odd components solely. The first condition is satisfied for any collinear anti-ferromagnetic SDW, while the second condition means that each half-period (without global phase shift) has mid-point mirror symmetry. Hence, the following expression has been finally adopted [11, 21]:

$$B(qx) = \sum_{n=1}^{N} h_{2n-1} \sin[\,(2n-1)\,qx\,]\,. \tag{2}$$

Symbols $h_{2n-1}$ denote amplitudes of subsequent harmonics. The symbol $q$ stands for the wave number of SDW, while the symbol $x$ denotes relative position of the resonant nucleus along propagation direction of the stationary SDW. The index $N$ enumerates maximum relevant harmonic. The argument $qx$ satisfies the following condition $0 \le qx \le 2\pi$ due to the periodicity of SDW. In fact, complete information is obtained having calculated expression (2) within the range $0 \le qx \le \pi/2$. Above range strictly applies to incommensurate conditions, but it could be used to commensurate SDW with $(2\pi)^{-1}qd$ being far away from the ratio of small non-zero integers, where the symbol $d$ denotes the smallest distance between various (all equivalent) resonant atoms (sites) in the direction of propagation of SDW. Amplitude of the first harmonic $h_1$ is by definition positive, as the absolute phase shift between SDW and crystal lattice is generally unobservable by the method used. The average amplitude $\langle B \rangle$ of SDW described by expression (2) equals zero, of course, while the mean squared amplitude of above SDW is expressed as $\langle B^2 \rangle = \frac{1}{2}\sum_{n=1}^{N} h_{2n-1}^2$ [11]. The last parameter is finite for any physical shape of SDW even for the infinite number of harmonics involved. One has to remember that spectrum is sensitive to $|B(qx)|$. Equation (2) applies to SDW of the reduced dimensionality as well provided some local method of observation is used. Amplitudes of subsequent harmonics have been fitted to the spectral shape. Remaining hyperfine parameters were used as common for all resonant sites. The electric quadrupole interaction was treated in the first order approximation as it is small compared to the magnetic interaction for all relevant cases. It was assumed that samples have neither magnetic nor crystallographic order on the macroscopic scale. Spectra were processed by means of the *GmfpHARM* program in the standard transmission integral approximation. The latter program belongs to the *MOSGRAF-2009* suite [22]. The field distribution as seen by resonant atoms could be calculated provided all relevant amplitudes of expression (2) had been determined. The latter distribution, i.e., probability density function $\rho(B) \ge 0$ takes on the form [21]:

$$\rho(B) = N_0^{-1} \sum_{L(B)} \frac{\partial}{\partial B} \left\{ \hat{B}^{-1} \left[ \| B(qx) \| \right] \right\} \text{ with } N_0 = \int_0^{B_{max}} dB \sum_{L(B)} \frac{\partial}{\partial B} \left\{ \hat{B}^{-1} \left[ \| B(qx) \| \right] \right\}. \tag{3}$$

The symbol $B_{max}$ stands for the maximum hyperfine field due to SDW. It has to be noted that operator $\hat{B}^{-1}$ is generally non-unique except within limited ranges of the primary argument $qx$. Hence, summation goes over all ranges of $B$ with the summation index $L(B)$ being dependent on the actual value of $B$. The number of ranges is always finite and accountable. It is sufficient to consider primary argument $qx$ from the range $0 \le qx \le \pi/2$. The secondary argument $B$ is non-negative and upper bound. A distribution described by equation (3) has singularities (singular maxima) at smooth extrema of the function described by expression (2). Singularity at maximum of the function (2) has sharp edge above singularity, while singularity at minimum of the function (2) has sharp edge below singularity. Maxima of the distribution occur also at smooth inflection points of the function (2). They are singular provided $\partial B / \partial(qx) = 0$ at this point. All above singularities are integrable (distribution is integrable and hence it is normalized), they are separated one from another, and one has finite number of such singularities. Singularities (maxima) might overlap for some SDW shapes. Therefore it is practical to approximate "exact" distribution given by equation (3) by the following expression removing singularities:

$$\rho_A(B) = \alpha^{-1} \left( \frac{1}{\Delta B} \right) \int_{B-\frac{1}{2}\Delta B}^{B+\frac{1}{2}\Delta B} dB' \, \rho(B') \text{ with } \alpha = \int_{\frac{1}{2}\Delta B}^{B_{max}-\frac{1}{2}\Delta B} dB \left\{ \left( \frac{1}{\Delta B} \right) \int_{B-\frac{1}{2}\Delta B}^{B+\frac{1}{2}\Delta B} dB' \, \rho(B') \right\} \text{ and}$$
$$\tfrac{1}{2}\Delta B \le B \le B_{max} - \tfrac{1}{2}\Delta B. \tag{4}$$

The last expression converges to equation (3) provided $\Delta B \to 0$. In fact, expression (4) generates histogram (with equal steps), the latter approximating distribution described by equation (3). Such histogram is adequate due to the limited resolution of the spectrum (experimental data) provided it is sufficiently dense. A distribution (probability density function) equals null for fields higher than $B_{max}$, of course. It is interesting to note that rectangular SDW leads to the distribution having character of the Dirac delta function, while triangular SDW leads to the distribution being completely flat. For purely sinusoidal SDW (all amplitudes equal zero except $h_1$) one gets $\rho(B) = [2/(\pi B_{max})] \, [1-(B/B_{max})^2]^{-1/2}$ with $B_{max} = h_1$. An integrable singularity appears for $B = B_{max}$. This singularity has sharp edge for fields above $B_{max}$. It has to be mentioned that the wave number of SDW is inaccessible experimentally to local methods like the Mössbauer spectroscopy. Expression (2) could be applied regardless of the angle between hyperfine field and propagation direction of SDW provided above angle remains constant. In general distributions described by equation (3) and subsequently by expression (4) are experimentally defined for the correlation between SDW and crystal lattice being random. In such case expression (2) is exact as well. Above random correlation conditions are satisfied for the most of cases with SDW being incommensurate. Finally, one has to observe that Mössbauer spectroscopy is one of the best methods to look upon details of the shape of SDW provided hyperfine fields are sufficiently large. Histograms generated by equation (4) are presented here in the form $W(B) \sim \rho_A(B)\Delta B$ (with constant and positive $\Delta B$) in order to conform to the standard presentation of the field distribution by Mössbauer spectroscopists. Each histogram $W(B)$ is normalized to unity. It has to be mentioned that spectrum due to single field (or several precisely defined fields) cannot be fitted within above model. The reason for that is as follows: such spectrum corresponds to ideal rectangular SDW, and for the latter one has logarithmically divergent sum of amplitudes.

## 4. Results and discussion

Figure 1 shows spectra obtained in the non-magnetic region and slightly above magnetic ordering temperature. Some onset of the magnetic order could be detected in the lowest temperature spectra of this region, as fits without hyperfine magnetic field are definitely of the poorer quality. On the other hand, the field is too small to be reliably included in the data processing. Spectra were fitted with either symmetric doublet or singlet. Singlet was used for $BaFe_2As_2$ compound as no discernible splitting exists above magnetic ordering temperature for this compound. Some slight preferential orientation could be seen in the case of doublets – particularly for $CaFe_2As_2$ compound with the largest quadrupole splitting. This orientation is too weak to have significant effect in the magnetically ordered region. All relevant parameters extracted from the data obtained in the non-magnetic and magnetic regions are summarized in Table I. The parameter $\Delta$ in Table I describes electric quadrupole interaction. For non-magnetic spectra it stands for the quadrupole splitting, while for magnetic spectra it stands for $\frac{1}{4}\left(\frac{c}{E_0}\right)eQ_eV_{zz}(3\cos^2\theta-1)$. Here the symbol $c$ stands for the speed of light in vacuum, the symbol $E_0$ stands for the energy of the resonant transition and the symbol $e$ stands for positive elementary charge. The symbol $Q_e$ denotes spectroscopic nuclear quadrupole moment in the excited nuclear state (positive for the first excited level in $^{57}$Fe), while the symbol $V_{zz}$ stands for the principal component of the electric field gradient (EFG) with the maximum absolute value. The angle $0 \le \theta \le \pi/2$ stands for the angle between above EFG component and the hyperfine field under assumption that EFG is axially symmetric. Axial symmetry is expected for the tetragonal phase, while the orthorhombic distortion is too small to have measurable effect on EFG. The EFG component $V_{zz}$ is expected to be positive here due to the local symmetry around iron atom, and to be oriented perpendicular to the tetragonal (orthorhombic) plane. Above statement is consistent with the observed sign of the very slight asymmetry of doublets in the non-magnetic region upon having taken into account the fact that absorber particles are composed of flat flakes likely to orient in the absorber plane. Tetragonal (orthorhombic) axis is perpendicular to the flake surface, of course. Hence, one can conclude that the hyperfine field is likely to be oriented closer to the orthorhombic plane than to the tetragonal (orthorhombic) axis as the angle $\theta$ exceeds magic angle. The quadrupole interaction parameter is not quoted for the non-magnetic spectrum of the $EuFe_2As_2$ compound obtained at 210 K, as it cannot be determined reliably due to the onset of magnetic order. One can use full diagonalization of the hyperfine Hamiltonian, but such procedure cannot be applied here as the electric quadrupole and magnetic dipole interactions, both, are very small and their parameters cannot be obtained reliably due to the limited resolution, i.e., due to finite line width. The quadrupole interaction in $BaFe_2As_2$ is too small to be observed without having sufficiently strong magnetic field. Spectral shift $S$ evolves with temperature mainly due to the second order Doppler shift (SOD). However, one can see slight decrease of the shift at transition from tetragonal to orthorhombic phase, i.e., an electron density on iron nucleus slightly increases while transforming tetragonal to orthorhombic phase. Similar effect, albeit of the opposite sign was observed in FeSe superconductor [2].

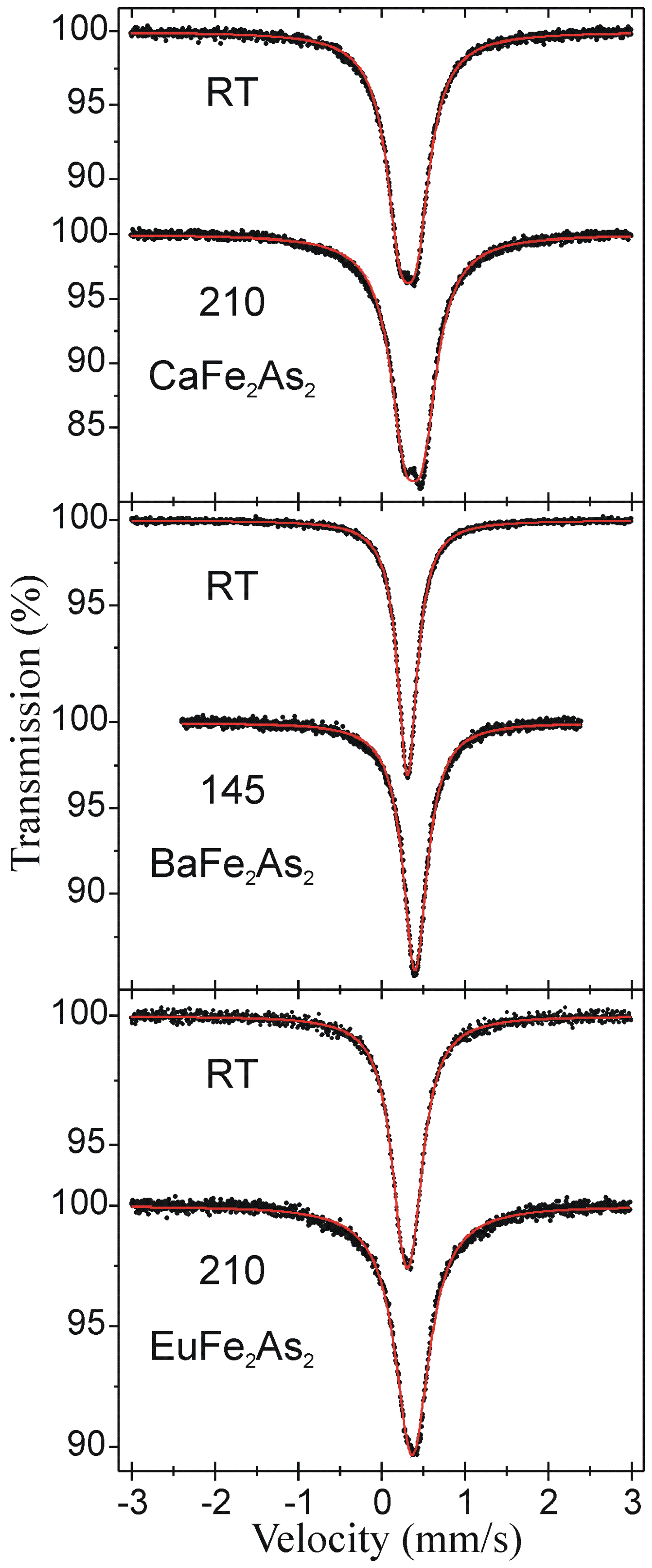

**Figure 1** Mössbauer spectra in the “non-magnetic” region.

## Table I

Essential parameters derived from the Mössbauer spectra. The symbol $T$ denotes temperature. The symbol $S$ denotes total shift (including second order Doppler shift) versus room temperature α-Fe, while the symbol $\Delta$ stands for the parameter describing electric quadrupole interaction (see text for details). Symbols $h_{2n-1}$ stand for the amplitudes of SDW harmonics. Values of the harmonic amplitudes quoted in brackets $<...>$ are obtained setting amplitudes of ten subsequent odd harmonics equal to the amplitude of the last quoted harmonic. Above common amplitude was used as variable. Such procedure was essential to reproduce quasi-rectangular shapes of SDW with high accuracy. Absorber line widths were used as 0.16 mm/s for Ca, 0.15 mm/s for Ba and 0.17 mm/s for Eu compounds in the magnetic region.

| $T$ (K) ±0.01 | $S$ (mm/s) ±0.001 | $\Delta$ (mm/s) ±0.002 | $h_1$ (T) ±0.01 | $h_3$ (T) ±0.02 | $h_5$ (T) ±0.03 | $h_7$ (T) ±0.03 | $h_9$ (T) ±0.03 | $h_{11}$ (T) ±0.03 | $h_{13}$ (T) ±0.03 |
|---|---|---|---|---|---|---|---|---|---|
| | | | | **$CaFe_2As_2$** | | | | | |
| RT | 0.434 | 0.194 | | | | | | | |
| 210 | 0.499 | 0.212 | | | | | | | |
| 180 | 0.517 | -0.044 | 5.39 | -2.05 | 1.05 | 0.12 | 0.23 | | |
| 170 | 0.516 | -0.078 | 7.17 | -1.64 | 0.10 | 0.21 | 0.24 | -0.06 | 0.16 |
| 165 | 0.510 | -0.094 | 8.01 | -1.15 | -0.24 | 0.06 | 0.31 | -0.02 | 0.08 |
| 160 | 0.511 | -0.096 | 8.63 | -0.71 | -0.34 | -0.12 | 0.30 | 0.01 | 0.10 |
| 155 | 0.508 | -0.098 | 9.10 | -0.35 | -0.32 | -0.20 | 0.25 | 0.03 | 0.11 |
| 145 | 0.514 | -0.094 | 9.74 | 0.13 | -0.26 | -0.32 | 0.18 | 0.01 | 0.17 |
| 120 | 0.531 | -0.090 | 10.71 | 0.99 | -0.17 | -0.35 | 0.23 | -0.29 | 0.28 |
| 80 | 0.550 | -0.092 | 11.56 | 1.65 | 0.23 | -0.30 | 0.07 | -0.27 | 0.19 |
| 4.2 | 0.566 | -0.092 | 12.41 | 2.45 | 0.80 | -0.02 | 0.46 | -0.30 | 0.03 |
| | | | | **$BaFe_2As_2$** | | | | | |
| RT | 0.426 | 0 | | | | | | | |
| 145 | 0.517 | 0 | | | | | | | |
| 141 | 0.518 | -0.004 | 1.39 | -1.18 | 0.83 | | | | |
| 139 | 0.519 | -0.014 | 1.70 | -1.42 | 0.95 | | | | |
| 137 | 0.517 | -0.028 | 3.31 | -0.92 | -0.45 | | | | |
| 135 | 0.515 | -0.028 | 4.41 | 1.08 | 0.11 | -0.46 | | | |
| 133 | 0.520 | -0.030 | 4.75 | 1.89 | 0.45 | 0.79 | | | |
| 131 | 0.520 | -0.030 | 5.01 | 1.92 | 0.62 | 0.71 | | | |
| 129 | 0.521 | -0.030 | 5.24 | 2.03 | 0.68 | 0.74 | 0.17 | | |
| 125 | 0.522 | -0.030 | 5.31 | 2.19 | 0.79 | 1.06 | 0.36 | 0.64 | |
| 105 | 0.536 | -0.032 | 5.95 | 2.66 | 0.88 | 1.26 | 0.44 | 0.72 | <0.25> |
| 80 | 0.540 | -0.034 | 6.68 | 1.97 | 0.94 | <0.42> | | | |
| 4.2 | 0.552 | -0.038 | 6.90 | 2.15 | 1.11 | <0.44> | | | |
| | | | | **$EuFe_2As_2$** | | | | | |
| RT | 0.425 | 0.118 | | | | | | | |
| 210 | 0.483 | | | | | | | | |
| 200 | 0.487 | -0.058 | 3.64 | -2.78 | 0.27 | | | | |
| 195 | 0.488 | -0.070 | 4.96 | -1.80 | 0.33 | 0.42 | 0.40 | | |
| 193 | 0.490 | -0.062 | 5.41 | -1.56 | 0.02 | 0.27 | 0.41 | | |
| 191 | 0.483 | -0.088 | 6.14 | -0.96 | -0.30 | -0.11 | 0.25 | | |
| 189 | 0.477 | -0.082 | 6.89 | -0.29 | -0.15 | -0.29 | 0.15 | | |
| 187 | 0.482 | -0.088 | 7.27 | 0.17 | -0.02 | -0.32 | 0.06 | | |
| 185 | 0.485 | -0.086 | 7.52 | 0.49 | 0.10 | -0.26 | 0.04 | | |
| 180 | 0.494 | -0.094 | 8.12 | 1.07 | 0.42 | -0.10 | 0.16 | | |
| 170 | 0.501 | -0.096 | 8.65 | 2.14 | 0.53 | -0.15 | 0.21 | 0.55 | |
| 145 | 0.520 | -0.104 | 8.86 | 3.90 | 1.21 | 1.93 | 0.49 | 0.44 | |
| 80 | 0.540 | -0.120 | 9.50 | 4.20 | 1.40 | 1.96 | 0.61 | 0.96 | 0.30 |
| 4.2 | 0.552 | -0.116 | 10.67 | 3.23 | 1.61 | 0.89 | 0.43 | 0.43 | <0.43> |

Figures 2-4 show spectra within temperature range with magnetic order. Each spectrum is accompanied by corresponding diagram showing SDW shape and by resulting histogram of the field distribution $W(B)$. Average fields $\langle |B| \rangle$ derived from histograms are shown for each histogram. In fact, such average field equals the average field calculated from the distribution of the absolute values of the hyperfine fields. Some common features could be recognized for all compounds investigated. For temperatures just below transition temperature to the magnetically ordered state SDW is composed of thin sheets with large volume of the crystal having very small amplitude of the spin density. However, the maximum amplitude is not very much smaller than at saturation (about 70 % of the saturation value). Upon lowering temperature magnetic sheets expand mainly at the base leading to the quasi-triangular shape, and finally quasi-rectangular shape occurs. Quasi-triangular behavior is definitely less pronounced for barium compound, and hence less harmonics suffice to describe spectrum. On the other hand, calcium compound does not develop fully quasi-rectangular shape even at liquid helium temperature. It is worth to note that SDW fully develops to the quasi-rectangular shape in much narrower temperature range for barium compound in comparison with calcium and europium compounds. Quality of fits is somewhat diminished in the quasi-rectangular region of SDW, as the latter spectra tend to be indistinguishable from spectra due to the unique field. However, one can be confident about the SDW model integrity due to the shape of the higher temperature spectra. Finally, one has to bear in mind that SDW amplitudes seen by resonant nuclei are re-scaled by the (almost) constant factor due to the core polarization. Hence, amplitudes in the absence of the atomic core and resulting magnetic moments cannot be simply estimated without detailed calculations of the iron electronic states all the way down to the deepest shells. The evolution of the SDW shape is seen in the relative units $qx$, and hence evolution in the absolute units of distance is additionally dependent on the variation of the wave number (wavelength) with the temperature.

Figure 5 shows comparison between the hyperfine field distribution derived from SDW and similar (normalized to unity) distribution obtained by means of the Hesse-Rübartsch (HR) method in the Lorentzian approximation and with the quadrupole interaction included to the first order approximation [23, 24]. Optimum smoothing filter and number of histogram points extended over the complete range of the hyperfine field have been applied in the case of HR analysis. The field range for HR data processing was set from zero to $B_{max}$ of the corresponding distribution due to SDW. Hence the scale of the hyperfine field ranges from $-\frac{1}{2}\Delta B_{\mathrm{HR}}$ till $B_{max} + \frac{1}{2}\Delta B_{\mathrm{HR}}$ in equal steps $\Delta B_{\mathrm{HR}}$ in the case of HR method. HR method is generally recognized as the best numerical approach to the spectra exhibiting field distribution of the unknown origin. Lower plots of Figure 5 show HR distributions for selected spectra, while upper plots show $W(B)$, the latter resulting from SDW for the same spectra. One could see that the average field $\langle |B| \rangle$ is reproduced rather accurately by the HR method, but the shape of the distribution is poorly reproduced - particularly in the flat regions. Gross maxima of the distributions due to SDW are reproduced approximately by the HR method. Fits are of the comparable quality, but the HR method does not give physical insight into the problem and one cannot go back from hyperfine field distribution to the shape of SDW, as the reciprocal transformation is not unique.

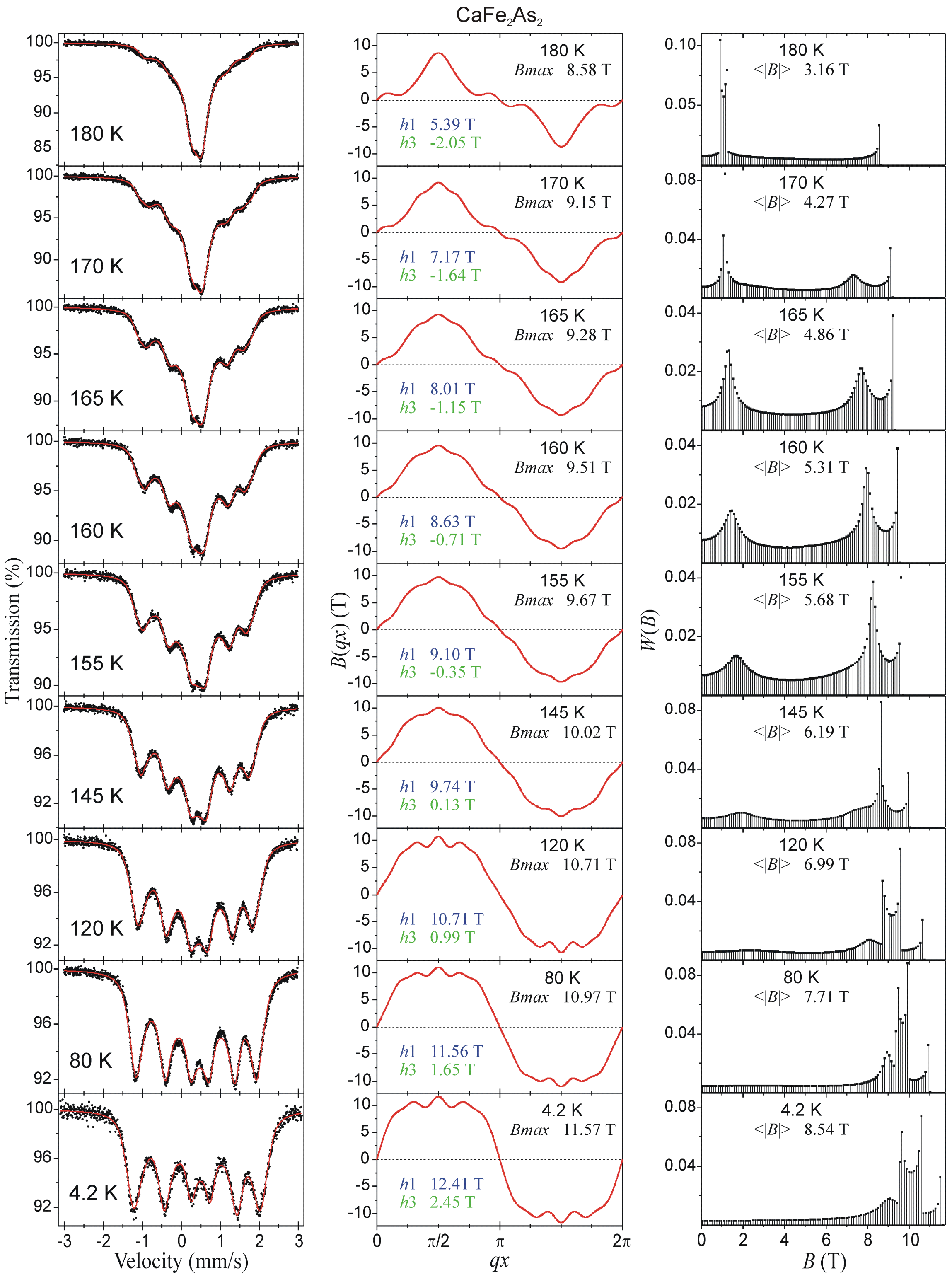

**Figure 2** Mössbauer spectra, SDW shapes and resulting hyperfine field distributions for $CaFe_2As_2$ in the magnetically ordered region versus temperature. Amplitudes of the first two harmonics $h_1$ and $h_3$ are listed on corresponding plots of SDW shape. Symbol $\langle|B|\rangle$ stands for the average of the respective field distribution, while $B_{max}$ denotes maximum value of SDW.

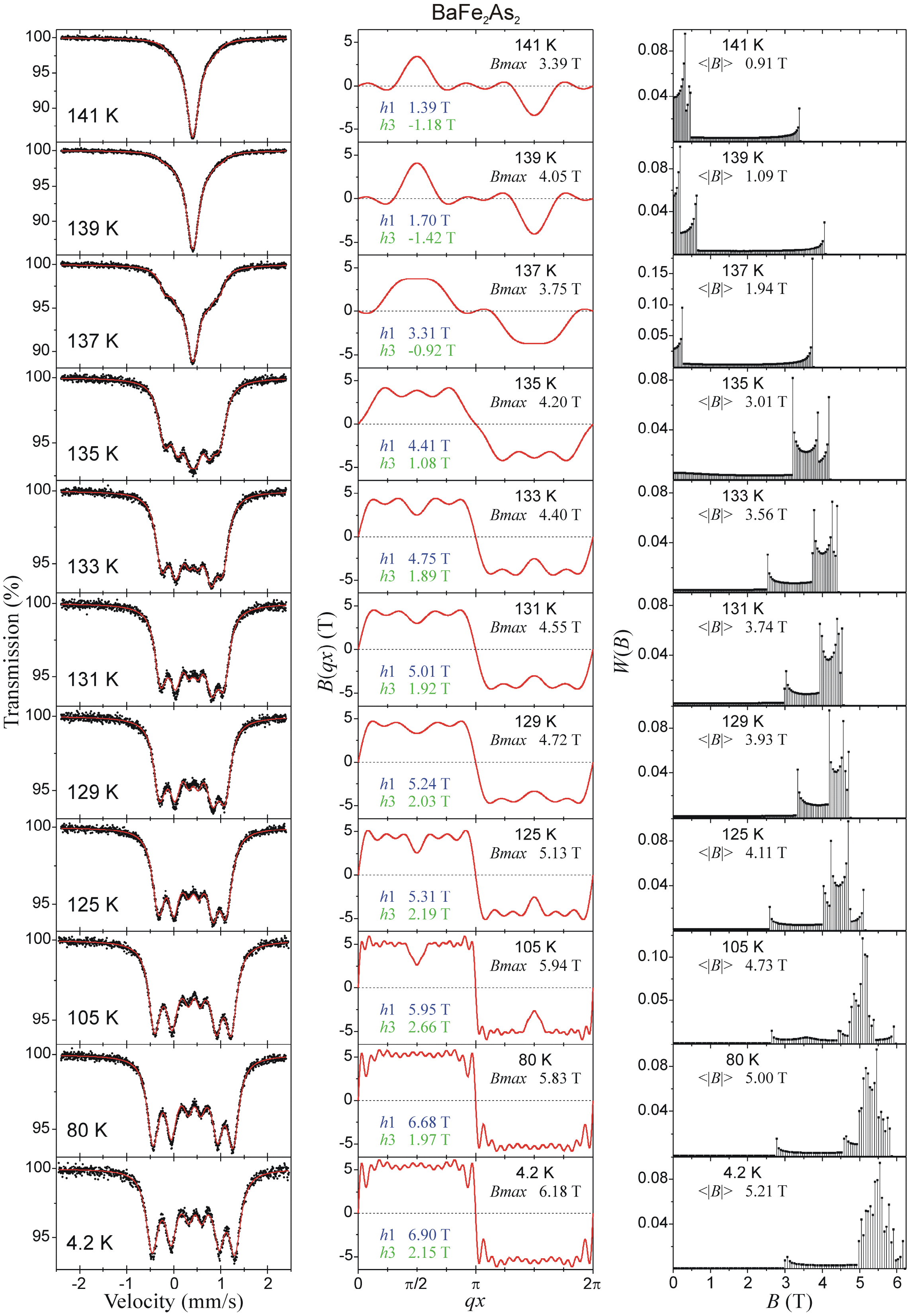

**Figure 3** Mössbauer spectra, SDW shapes and resulting hyperfine field distributions for $BaFe_2As_2$ in the magnetically ordered region versus temperature.

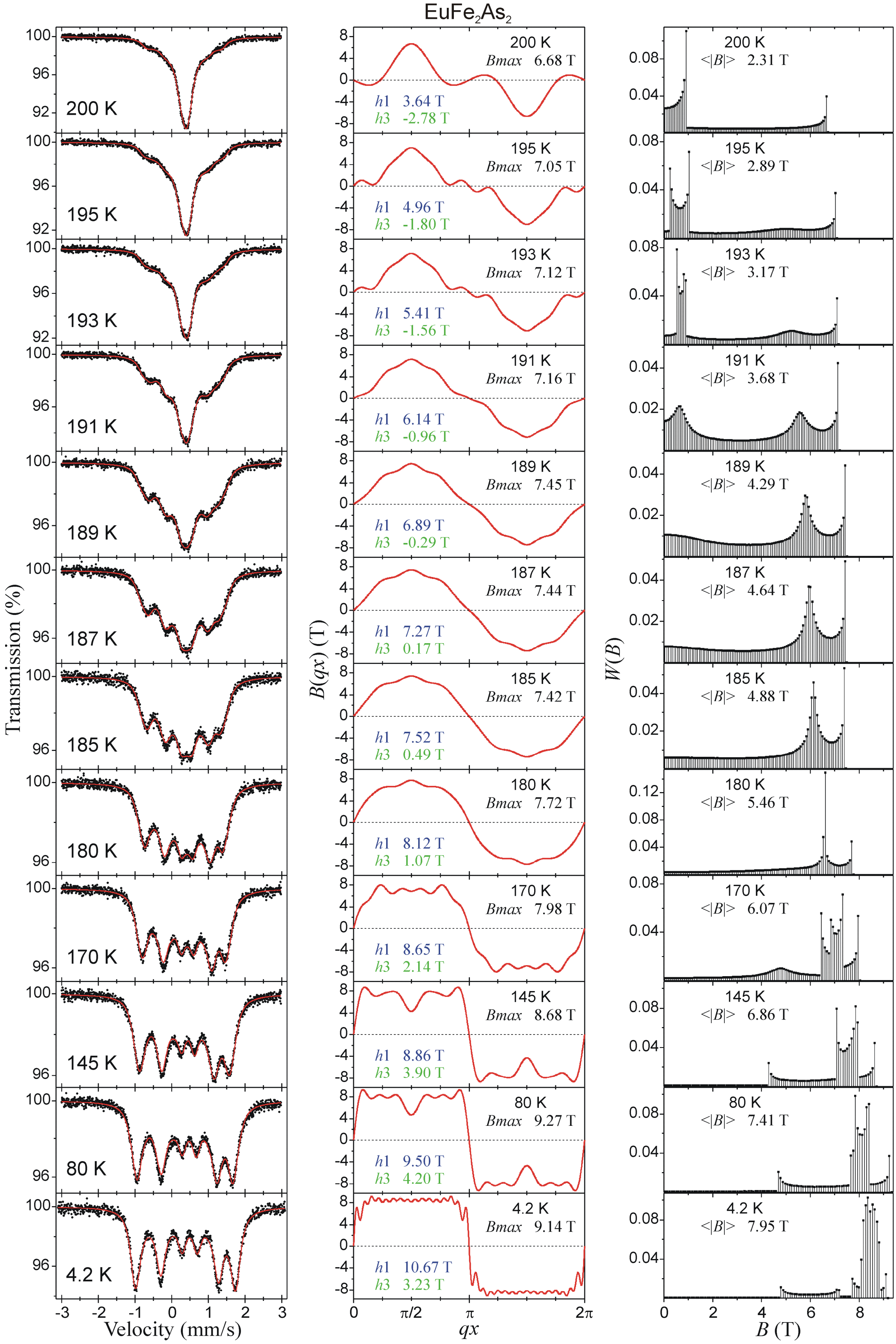

**Figure 4** Mössbauer spectra, SDW shapes and resulting hyperfine field distributions for $EuFe_2As_2$ in the magnetically ordered region versus temperature.

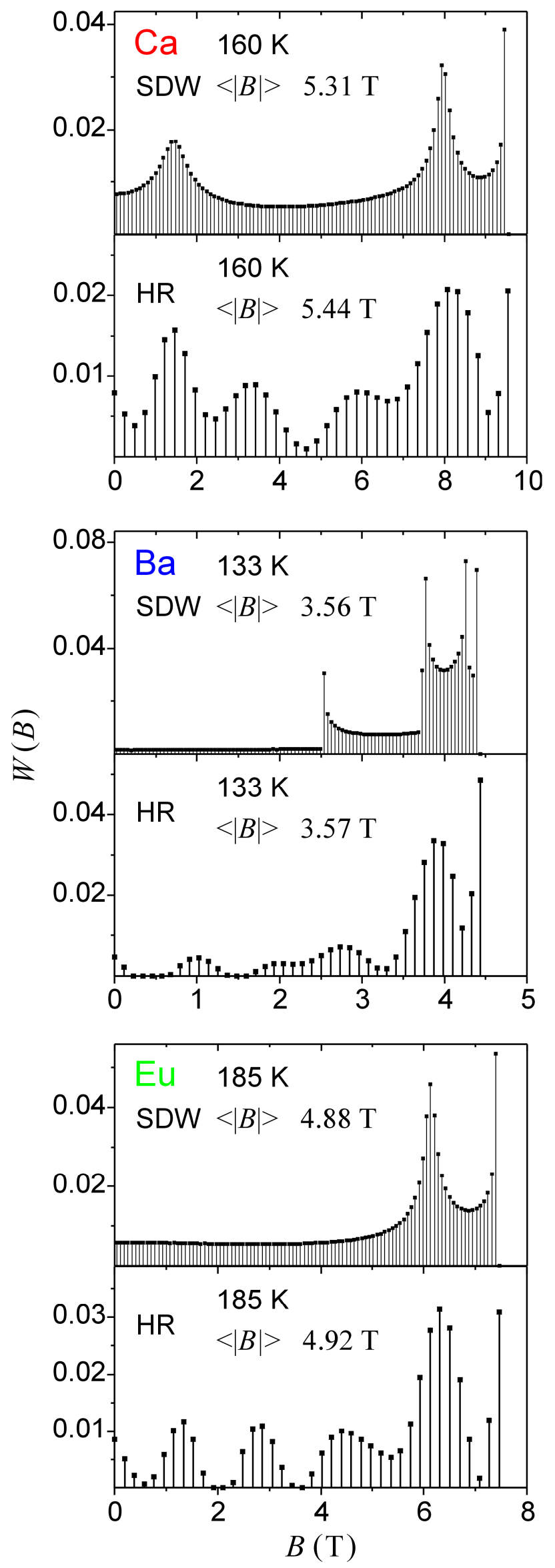

**Figure 5** Comparison of the hyperfine field distribution resulting from SDW shape with the corresponding field distribution yielded by the HR method for selected spectra.

Figure 6 shows $\sqrt{\langle B^2 \rangle}$ versus temperature for all compounds investigated. This parameter is in principle proportional to the square root from the intensity of the magnetic Bragg reflection in the neutron scattering – provided that other electronic magnetic moments than due to SDW are absent and $q$ remains fairly constant [11]. Hence, the parameter $\sqrt{\langle B^2 \rangle}$ is proportional to the value of the electronic magnetic moment per unit volume. It is evident that magnetic ordering of europium at low temperature (at about 19 K) has no effect on the magnetism of 3d electrons.

It seems that the parameter $\sqrt{\langle B^2 \rangle}$ behaves versus temperature $T$ like the power function of the following type:

$$\sqrt{\langle B^2\rangle} = B_0\left(1-\frac{T}{T_c}\right)^{\alpha} \text{ for } 0 \le \frac{T}{T_c} \le x_0 \text{ and } \sqrt{\langle B^2\rangle} = B_0\, A\left(\frac{T}{T_c}\right)^{-\beta} \text{ for } \frac{T}{T_c} \ge x_0$$

$$\text{with } x_0 = \frac{\beta}{\alpha_0+\beta} \text{ and } A = x_0^{\beta}\,(1-x_0)^{\alpha_0}\,. \tag{5}$$

The symbol $B_0 > 0$ denotes saturation value of the ground state, while the symbol $T_c > 0$ denotes transition temperature to the high-temperature state. Low temperature (critical) exponent satisfies the following condition $0 < \alpha < 1$. The high temperature exponent $\beta$ is also positive and usually one can expect $\beta > 1$. It is hard to expect that the low temperature exponent remains constant within the whole applicable temperature range. Improved fits could be obtained setting $\alpha = \alpha_0[\,(1-\gamma)+\gamma\,(T/T_0)\,]$ for $0 \le T < T_0$ and $\alpha = \alpha_0$ for $T \ge T_0$, where the following conditions are satisfied $0 < \alpha_0 < 1$ and $0 \le \gamma < 1$. The temperature $T_0$ is defined as $T_0 = x_0 T_c$, while the parameter $\alpha_0$ approximates low temperature critical exponent. It is assumed here that the following condition is satisfied $(T_c - T_0)/T_c << 1$. It is unlikely that the high temperature tail extending till the transition (the latter having some hysteresis) from the low temperature orthorhombic phase to the magnetically disordered high temperature tetragonal phase is due to the critical fluctuations as it is too large, i.e., it represents relatively too strong field. On the other hand, critical fluctuations lead to the more or less homogeneous broadening of the Mössbauer spectrum, the latter being inconsistent with the observed spectrum shape within this temperature region. It rather seems that this tail is due to SDW having short coherence length, and hence invisible by the diffraction based methods like coherent neutron scattering. On the other hand, local methods like the Mössbauer spectroscopy are completely insensitive to the time independent incoherence of SDW. Therefore $\beta > 1$ is expected in contrast to the case of critical fluctuations. Hence, for coherent neutron scattering one is likely to see some quantity proportional to the following function:

$$\sqrt{\langle B^2\rangle_C} = B_0\left(1-\frac{T}{T_c}\right)^{\alpha} \text{ for } 0 \le \frac{T}{T_c} \le 1 \text{ and } \sqrt{\langle B^2\rangle_C} = 0 \text{ for } \frac{T}{T_c} \ge 1. \tag{6}$$

Equation (5) yields the same result as equation (6) for $T/T_c \le x_0$. Functions described by above equations bifurcate at temperature $T_0 < T_c$, i.e., below temperature $T_0$ the incoherent contribution vanishes, while above temperature $T_c$ coherent contribution vanishes. The field at temperature $T_0$ takes on the value $B_F = B_0\left(\frac{\alpha_0}{\alpha_0+\beta}\right)^{\alpha_0}$. Coherent and incoherent contributions equal each other at temperature $T_t = x_t T_c$, where $T_0 < T_t < T_c$. The parameter $x_t$ is solution of the equation $Ax_t^{-\beta} - 2(1-x_t)^{\alpha_0} = 0$ from the range $x_0 < x_t < 1$. The total field $B_t$ at temperature $T_t$ is composed of coherent and incoherent contributions in equal amounts. Above discrepancy between coherent and local methods has been already mentioned by Bonville *et al.* [11]. It is interesting to note as well that equation (5) yields the following asymptotic form close to the ground state:

$$\lim\left(\sqrt{\langle B^2\rangle}\right)_{T/T_c\to+0} =$$

$$= B_0\left[1-\alpha_0(1-\gamma)\left(\frac{T}{T_c}\right)-\tfrac{1}{2}\left(\alpha_0(1-\gamma)[1-\alpha_0(1-\gamma)] + \frac{2\alpha_0\gamma T_c}{T_0}\right)\left(\frac{T}{T_c}\right)^2+\ldots\right]. \quad (7)$$

The linear behavior of equation (7) versus temperature close to the ground state could be seen unless $\alpha_0(1-\gamma)\approx 0$. Essential results of fits are summarized in Table II. The linear dependence close to the ground state is clearly seen for $CaFe_2As_2$, while for remaining compounds the condition $\alpha_0(1-\gamma)\approx 0$ is approximately satisfied as SDW close to the ground state has almost rectangular shape. For these compounds quasi-anti-ferromagnetic conditions are reached in contrast to $CaFe_2As_2$. Hence, the itinerant character of magnetic order is pronounced at most for $CaFe_2As_2$. It is interesting to note that the critical exponent $\alpha_0$ is quite close to the critical exponent (~0.125) of the anti-ferromagnetic system having one dimension in the spin space and two dimensions in the real space. Uni-dimensional spin space is consistent with the SDW being planar wave with definite spin polarization, i.e., for unique direction of the spin orientation. Two dimensions in the real space mean that the exchange interaction leading to the magnetic order is predominantly confined to (orthorhombic) planes and it seems that coupling between planes weakens, while going from A=Ca through Eu to Ba. The same universality class $(1,2)$ with uni-dimensional electronic spin space and two-dimensional real space has been found for $EuFe_2As_2$ by Raffius *at al.* [25] as far as the 3d magnetic order is concerned. Note that the exponent $\beta$ increases drastically, while going from Ca through Eu to Ba. It means that magnetic order is more and more confined to the orthorhombic plane when calcium is replaced by europium and subsequently by barium. Such behavior correlates with the lattice period along orthorhombic axis. This lattice constant is c=11.725 Å for $CaFe_2As_2$ [18], c=12.085 Å for $EuFe_2As_2$ [19] and c=13.017 Å for $BaFe_2As_2$ [6] in the tetragonal phase at room temperature. The presence of the magnetic tail could be due to the co-existence of the magnetically disordered tetragonal phase and magnetically ordered orthorhombic phase exhibiting small magnetic hyperfine field. However, this interpretation does not seem likely for the following reasons. First of all no irreversibility of the parameter $\sqrt{\langle B^2\rangle}$ versus temperature was found near transition temperature. Furthermore coherent regions of the parent/daughter phase are likely to have size comparable with the size of the crystallite or domain in the case of large single crystal. This size exceeds typical coherence length of the neutron scattering method, while the tail is still invisible by the neutron scattering method [26]. The nuclear magnetic resonance (NMR) results obtained by means of the $^{75}$As nuclear probe are less conclusive due to the complex mechanism leading to the transferred hyperfine field on the As nucleus [27-29].

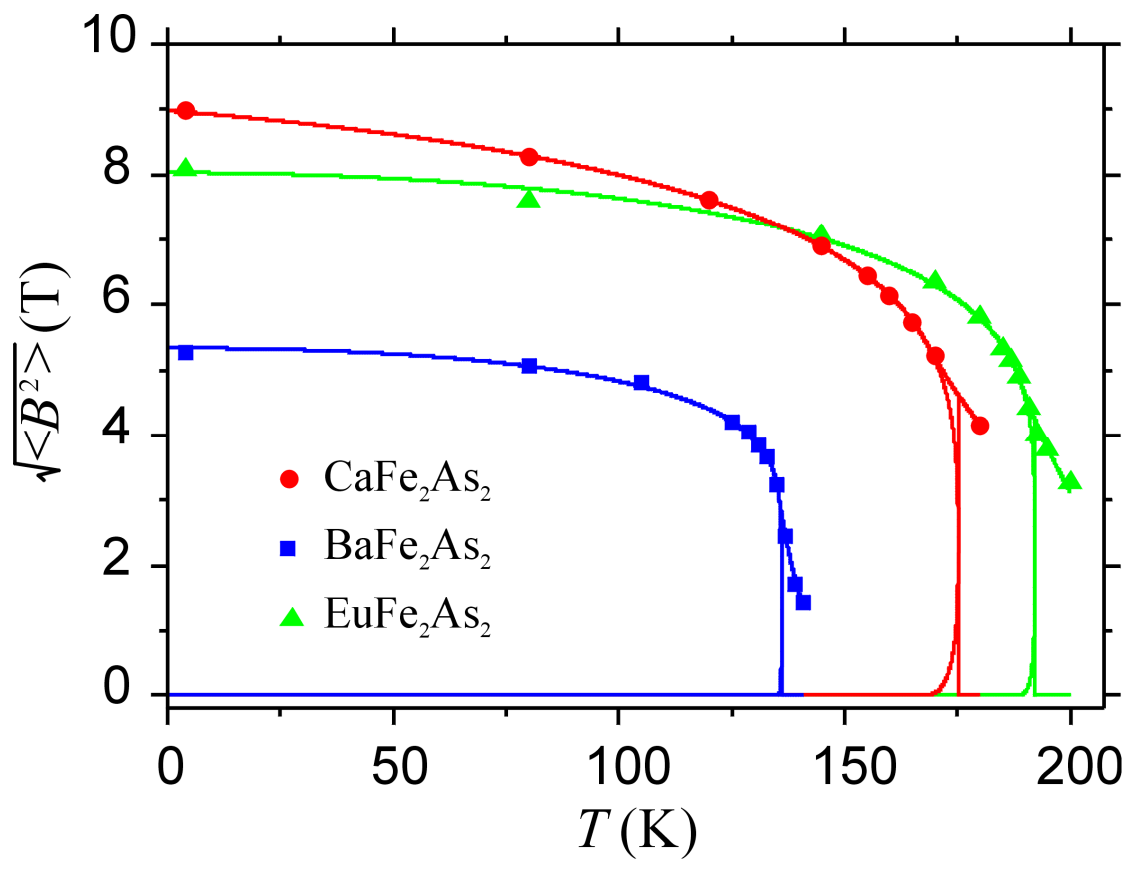

**Figure 6** Plot of square root from the mean squared amplitude of SDW $\sqrt{\langle B^2\rangle}$ versus temperature. Solid lines represent total, coherent and incoherent contributions as described in the text. Error bars for both field and temperature are smaller than the symbol size.

**Table II**

Essential parameters describing evolution of $\sqrt{\langle B^2 \rangle}$ versus temperature. Note, that parameter $B_0 A$ stands for the incoherent field at temperature $T_c$. Fitted parameters are $B_0$, $T_c$, $\alpha_0$, $\gamma$ and $\beta$. Remaining parameters are derived basing on the values of the fitted parameters. Errors for parameters $B_t$ and $T_t$ have not been estimated as these parameters depend on the numerical solution of the non-linear equation – see text for details. Transition temperatures $T_c$ are close to those reported in the literature, i.e., 171 K for $CaFe_2As_2$ [4], 140 K for $BaFe_2As_2$ [6] and 190 K for $EuFe_2As_2$ [5].

| **Compound** | **$CaFe_2As_2$** | **$BaFe_2As_2$** | **$EuFe_2As_2$** |
|---|---|---|---|
| $B_0$ (T) | 8.98(2) | 5.34(1) | 8.03(2) |
| $B_F$ (T) | 5.4(1) | 3.1(1) | 4.8(1) |
| $B_t$ (T) | 4.62 | 2.81 | 4.25 |
| $B_0 A$ (T) | 4.62(3) | 2.81(2) | 4.25(2) |
| $T_c$ (K) | 175.3(3) | 136.0(1) | 192.1(1) |
| $T_t$ (K) | 175.2 | 136.0 | 192.1 |
| $T_0$ (K) | 168.3(4) | 135.3(1) | 189.1(1) |
| $\alpha_0$ | 0.158(2) | 0.102(1) | 0.124(1) |
| $\gamma$ | 0.3(1) | 1.0(1) | 0.9(1) |
| $\beta$ | 3.8(1) | 20.4(4) | 7.6(1) |

It seems that similar kind of magnetic order occurs in other parents or under-doped compounds leading to the iron-based superconductivity provided they behave like metals. In particular Mössbauer iron spectra of the 1111 family are quite similar for the magnetically ordered phases [1, 9, 10, 17]. Some of them were interpreted as due to several (or single) precisely defined fields, but such treatment does not seem justified, while looking for the evolution of the spectrum shape with temperature. Another explanation was given interpreting spectra within framework of the spin fluctuations, i.e., as the relaxation spectra [10]. Such fluctuations are unlikely to be observable on the so long time scale as the time scale of the hyperfine interactions. Overall density of electronic spins is high, magnetic order develops at rather high temperatures and the system is metallic. Hence, the relaxation has to occur on much shorter time scales leading to the semi-classical average field in the conduction band. Relaxation is mainly due to the spin-spin interactions with very small lattice contribution as the system has itinerant electrons. Hence, there is no way to explain dependence of the relaxation time on temperature. Even rare earths (if present as constituents) develop rather hyperfine field due to long range order instead of coupling between nucleus and the local 4f shell despite the fact that 4f electrons are weakly coupled to the conduction band [10, 30, 31]. It seems that sample inhomogeneity leading subsequently to magnetic clusters and hyperfine field distribution could be ruled out at least for parent compounds, as the latter are highly ordered structures from the crystallographic point of view. Other mechanisms of magnetic ordering like e.g. spin glass formation are even less probable for compounds in question due to the strongly layered structure.

## 5. Conclusions

Study of the evolution of the Mössbauer spectra versus temperature for $AFe_2As_2$ (A=Ca, Ba, Eu) parent compounds of the 122 family of iron-based superconductors leads to the conclusion that 3d magnetism of these compounds is due to development of SDW below the transition temperature. The shape of SDW evolves significantly with temperature forming narrow sheets of significant spin density at the onset of the transition. The magnetic sheets broaden subsequently passing in some cases through the semi-triangular shape of SDW. Finally, quasi-rectangular shape is formed and spectra are almost the same as for single precisely defined field, the latter being characteristic of the classical anti-ferromagnet (and ferromagnet, too). It seems that such behavior is common feature for all parents and related under-doped compounds e.g. also for the 1111 family. No measurable charge density waves (CDW) were found at least in the bulk of the material, the latter contributing most of the signal in the transmission Mössbauer spectroscopy.

The critical exponent $\alpha_0$ cannot be determined very accurately due to the presence of the tail above magnetic transition and in fact it is rather underestimated. However, its value indicates rather with any doubt that these compounds belong to the universality class $(1,2)$ as far as the magnetic order is considered. The tail is likely to be due to the presence of incoherent SDW in relatively narrow temperature range around transition point.

There is no measurable contribution to the magnetic hyperfine field on iron due to the magnetic ordering of the europium ions. Recently a small transferred field on iron due to the europium ordering was found in $EuFe_2(As_{1-x}P_x)_2$ even in the superconducting state with diamagnetic iron accompanied by the generation of the significant electric quadrupole interaction on iron owing to the replacement of As by P [32].

Formation of the magnetically ordered sheets separated by almost non-magnetic layers is prerequisite to the formation of the Fulde-Ferrell-Larkin-Ovchinnikov (FFLO) phase [33, 34]. However, non-magnetic regions do not exhibit superconductivity for the compounds in question, and upon lowering temperature the system orders in the itinerant anti-ferromagnetic fashion – not necessarily commensurate. It seems that SDW period in the real space is too small to allow formation of the FFLO phase, i.e., the wave number $q$ is too high for the systems investigated.

### Acknowledgments

Dr. Jakub Cieślak from the Faculty of Physics and Applied Computer Science, AGH University of Science and Technology, Kraków, Poland is thanked for informative discussions.